 \documentclass[]{aa}  
\usepackage{graphicx}
\usepackage{url}
\usepackage{lscape}
\usepackage{rotating}
\usepackage{natbib}
\usepackage{txfonts}
\begin{document}
   \title{Chemical composition of the Taurus-Auriga association\thanks{Based on observations collected at Telescopio Nazionale Galileo, Canary Islands (Spain). 
   Program AOT16/07B}
   }


   \author{Valentina D'Orazi
          \inst{1}
        \and
	Katia Biazzo\inst{2}  
	\and
          Sofia Randich\inst{2}
	  }
	  
 \offprints{V. D'Orazi}

  \institute{INAF $-$ Osservatorio Astronomico di Padova, Vicolo dell'Osservatorio 5, I-3522, Padova, Italy.\\
              \email{valentina.dorazi@oapd.inaf.it}
         \and
             INAF $-$ Osservatorio Astrofisico di Arcetri, Largo E. Fermi 5,
	     I-50125, Firenze, Italy\\
             }
 
   \date{Received: Accepted:}
  
 
  \abstract
   {The Taurus-Auriga association is perhaps the most famous prototype of a low-mass star forming region, surveyed at almost all
   wavelengths. Unfortunately, like several other young clusters/associations, this T association lacks an extensive 
   abundance analysis determination.}
   {We present a high-resolution spectroscopic study of seven low-mass members of Taurus-Auriga, including both weak-lined and
   classical T Tauri stars designed to help robustly determine their metallicity. }
   {After correcting for spectral veiling, we performed equivalent width and spectral
   synthesis analyses 
   using the GAIA set of model atmospheres and the 2002 version of the code MOOG.}
   {We find a solar metallicity, obtaining 
   a mean value of [Fe/H]=$-$0.01$\pm$0.05. The $\alpha$-element Si and the Fe-peak one Ni confirm a solar composition.
  Our work shows that the dispersion among members is well within the observational errors
   at variance with previous claims. As in other star forming regions, no metal-rich members are found, reinforcing the idea that old planet-host
stars form in the inner part of the Galactic disc and subsequently migrate.}
   {}

  \titlerunning{Abundances in Taurus-Auriga}
  \authorrunning{D'Orazi, Biazzo \& Randich}             
   \maketitle

\section{Introduction}\label{introduction}

The Taurus-Auriga (Tau-Aur) association is the nearest large star-forming region (SFR), with a close 
distance of $d$=140 pc (e.g., Kenyon et al. 1994).
Given its lack of massive O/B stars, this T association in the past 50 years has become a standard region to study the low-mass star 
formation processes.
Tau-Aur is widely spread across the Northern sky ($\sim$100 square degrees, Scelsi et al. 2007a) and contains
hundreds of low-mass members; Rebull et al. (2010) found 148 new candidate members, of which 34 were confirmed 
by spectroscopic follow-up.
The bulk of stars presents an average age of $\sim$1 Myr (e.g., Brice\~{n}o et al. 1999).

The whole region has been observed at almost all wavelengths, from infrared to X-rays 
(e.g., Itoh et al. 1996; Brice\~no et al. 1999; Davis et al. 2008; Luhman et al. 2010). 
Kenyon \& Hartmann (1995, hereafter KH95) provided a thorough investigation of the stellar population, from 
class 0 (proto-stars) to class II/III (i.e. classical and weak-lined T Tauri stars, respectively -hereafter CTTs and WTTs). 
They analysed optical and infrared photometric observations and presented colour-magnitude and colour-colour diagrams, luminosity and 
mass functions, along with 
information on the near-infrared excess and accretion properties.
Lithium abundance studies of this association have been also performed over the past decade, such as, e.g., Basri et al. (1991), Magazz\'u et al. (1992), 
Mart\`{i}n et al. (1994), and Sestito et al. (2008), who presented a (re)assessment of Li~{\sc i} abundances in WTTs and CTTs. 
G\"{u}del et al.~\cite{gud} carried out an extensive X-ray survey (XMM-{\it Newton Survey of the Taurus Molecular Cloud} --XEST) 
covering an area of $\sim$ 5 square degrees and concentrating mainly on the higher stellar density regions.

Whereas all these studies have helped to characterize the stellar population, 
Tau-Aur shares with other nearby SFRs the lack of an 
accurate abundance analysis.

The determination of chemical composition of SFRs is instead critically important to a variety 
of astrophysical issues, in both planetary and stellar contexts, as we previously discussed in our pilot project
focusing on the Orion association (D'Orazi et al. 2009, hereafter D09). At variance with OB associations, whose
abundances can provide an independent test of the so-called {\it triggered} star formation scenario (e.g., Blaauw 1991) and 
indicate whether there is local enrichment, we would expect to measure a very homogeneous composition for T associations.

As is well known, giant planets are preferentially found around metal-rich main sequence stars 
(Santos et al. 2004, and reference therein). 
The natural question hence arises as to whether in the first 
epochs of planetary formation (discs of T Tauri stars are commonly accepted as planet birthplaces) 
the probability of 
hosting a giant planet depends on the star's metallicity.

On the other hand, and very interestingly, several studies (e.g., Luhman  2004) 
have found significant differences between the initial mass function (IMF) derived for the Tau-Aur and those derived for denser systems 
containing massive members, e.g., the Trapezium, with the former containing a too small number of brown dwarfs  
according to the standard IMF (Brice\~no et al. 2002) and a surplus of late-K and early-M stars
(Luhman et al. 2009).
These differences can be attributed to different conditions of the environment where the stars were formed. 
Some studies have claimed that the IMF slope may depend on metallicity, in the sense that metal-rich environments tend to produce more low-mass stars 
(M $< 0.7 \rm M_{\odot}$) than metal-poor systems (Larson 2005; da Rio et al. 2009 and references therein).   
The obvious question is: could metallicity play a role in the lower fraction of very low-mass 
stars and brown dwarfs detected in Taurus with respect to Trapezium (Brice\~no et al. 2002)?
More generally, we may ask whether there is a difference between chemical compositions of 
low-mass and high-mass SFRs or, whether, SFRs represent a chemically homogeneous class of objects? \\


Available information on Tau-Aur abundances comes from the 
previous works by Padgett (1996) and Santos et al. (2008). 
First, Padgett (1996) analysed eight WTT members
and derived a solar composition, i.e. an 
average value of [Fe/H]=0.01$\pm$0.13 (rms), with a significant scatter  
among members ranging from $-$0.11 to +0.22. 
Santos et al. (2008) presented the iron abundances of three Taurus members: 
in this case, the spread in metallicity is also large, varying from $-$0.18 to +0.05 
with a mean value of [Fe/H]=$-$0.07$\pm$0.12 (rms). 
Given the quite large star-to-star variation obtained by both studies, a new accurate abundance analysis 
for this association is warranted, 
to minimize the impact of the observational uncertainties and ascertain whether the metallicity scatter
among members is indeed real or not. 

In a different framework, Scelsi et al. (2007b) obtained
coronal abundances for a sample of X-ray bright members of Taurus, 
finding that the iron abundance in the corona is
significantly (by a factor of $\sim 5$) lower than the solar photospheric value.
More generally, their coronal abundances revealed a pattern
in agreement with those obtained in
previous studies.
Specifically, X--ray observations of young and/or active
stars have unveiled a so--called FIP 
(first ionization potential)-dependent abundance trend:
low FIP elements (FIP$\leq$10 eV, like iron, silicon, nickel) are under-abundant in the corona
with respect to the photospheric values 
(the so-called inverse-FIP effect, for further details see Brinckman 
et al. 2001; Audard et al. 2003; Scelsi et al. 2007b). 
As stressed by Scelsi et al., however, this FIP-related pattern for Tau-Aur, as well as for other young stars, was obtained by comparing the coronal abundances to the solar photospheric values, rather
than the stellar photospheric abundances. Different or no FIP effects
were indeed found when considering the correct stellar photospheric abundances 
(e.g., Sanz-Forcada et al. 2004; Maggio et al. 2007).
Determining the photospheric abundances of Tau-Aur members with
measured coronal abundances thus appears critical to confirm whether
a FIP-related effect is present in these stars or not.

In this work, we present a high-resolution spectroscopic study based on seven low-mass members of the 
Tau-Aur association. 
The sample was selected specifically to determine accurate abundances, by discarding, e.g., fast rotators, strong accretors, and/or binary stars. 
Most importantly, this is one of the first sample that also includes T Tauri stars
that have not yet dissipated their circumstellar discs and reveal moderate accretion from the surrounding material. 
Details of the observations and data reduction are given in Section~\ref{obstau};  
Section~\ref{analysistau} describes the estimate of and correction for spectral veiling, along with the abundance analysis procedure. 
The results and the scientific implications are illustrated in Sections~\ref{results} and~\ref{discu}, while 
in Section~\ref{summary} a conclusive summary is given.  
\section{Observations and data reduction}\label{obstau}
Our sample consists of seven low-mass members of the Tau-Aur 
association, of which two are CTTs and five WTT stars, respectively.  The sample was selected from the 
large database provided in XEST by G\"{u}del et al. (2007). 

Single K-type members were chosen with projected rotational velocities (v$sin$i) not larger than $\sim$30 km s$^{-1}$ and a 
mass accretion rate below log~\.{M}=$-$7.5 M$_{\odot}$~yr$^{-1}$. 
For stars rotating faster than 30 km~s$^{-1}$ the abundance analysis becomes infeasible, 
because of rotational blending. The upper limit to the accretion rate allowed by including in the sample also CTTs  and, at the same time, discarding
extremely strong accretors, for which abundance analysis is no longer possible. Basic properties of our sample are given in 
Table~\ref{logtau}, where we list the star name, the observation date, and the exposure times in Columns 1, 2, 3, respectively; 
signal-to-noise ratio (S/N) per pixel measured around the Li~{\sc i} line at 6708 \AA~
in the co-added spectra are reported in Column 4.
The $V$ magnitudes and spectral-types are listed 
in Columns 5 and 6, respectively, while the initial effective temperature (T$_{\rm eff}$)
from G\"{u}del et al. (2007) is given in Column 7. 
Finally, in Column 8 we show the classification as CTT (Class II) or WTT (Class III) stars, while projected rotational 
velocities from XEST catalogue are in Column 9.
  
\begin{figure*}
\begin{center}
\includegraphics[width=12cm]{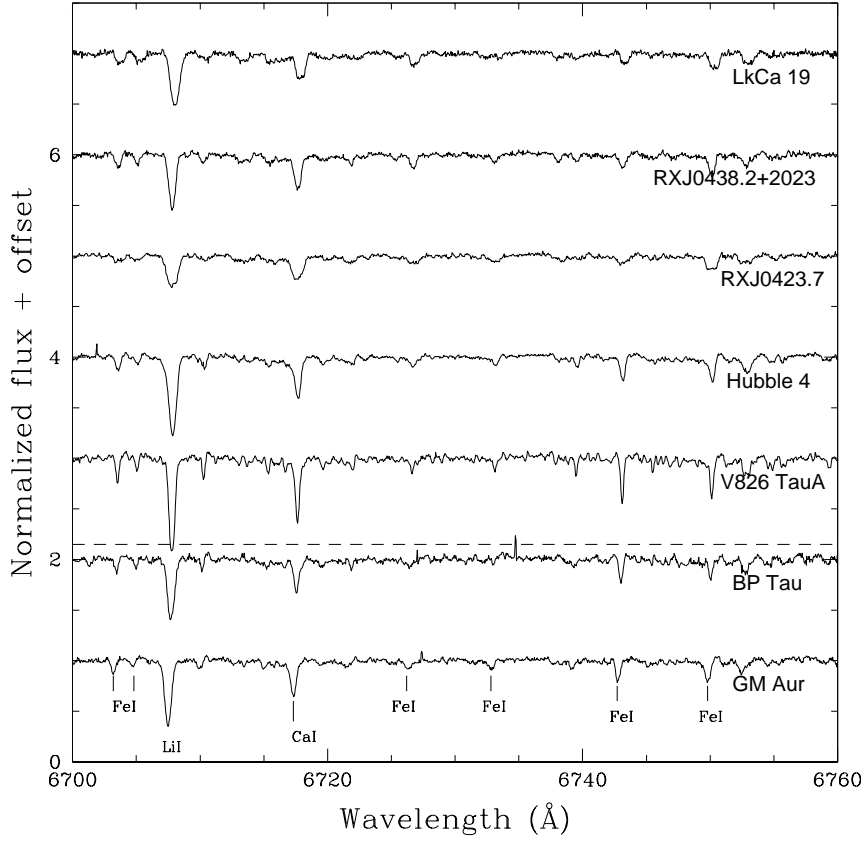}
\caption{Portion of the spectra of our seven sample stars between 6700~\AA~and 6760~\AA; the dashed line separates the WTTs from the two CTTs
(GM Aur and BP Tau).}
\label{spectratau}
\end{center}
\end{figure*}

The observations were carried out in service mode at the Telescopio Nazionale Galileo (Roque de los Muchachos, Canary Islands) with the high-resolution spectrograph SARG,
during seven nights of observations between August and December 2007. 
We used the 61 \AA~mm$^{-1}$ CD3 
yellow grism ($\lambda$$_{\rm blaze}$=5890 \AA), which resulted in a spectral
coverage between 4620 \AA~ and 7920 \AA~, along with a 1.6$^{"}$ slit, providing a resolution R $\sim$29000. 
As we discuss below, we also acquired a solar spectrum with the same instrumental set-up, which permits us to carry out 
a differential abundance analysis (see Sect.~\ref{analysistau}) with respect
to the Sun.
Data reduction was performed within the {\sc echelle} context in MIDAS following the standard procedures of bias subtraction, 
flat-field correction, order definition and
extraction, sky subtraction, and wavelength calibration. 
Spectra of the same star acquired 
in different exposures were then co-added, after checking that
there were no variations in radial velocity. 
The spectra of seven of our sample stars are shown in Figure~\ref{spectratau} in the 
spectral window 
between 6700 and 6760 $\AA$, where several Fe~{\sc i} features are present.

\setcounter{table}{0}
\begin{table*}
\begin{center}
\caption{Log of observations and stellar properties.}
\begin{tabular}{lcccccccr}
\hline\hline
Star & Obs.Date & Exp.time  & S/N$_{6708 \AA}$ & V      & Sp.Type & T$_{\rm
eff}$ & Class & v$sin$i$_{\rm lit.}$\\
     &  (yy-mm-dd)    & (sec)    &                     &  (mag) &         &   (K)         &       & (km~s$^{-1}$)\\
\hline     
GM Aur &   2007-09-08 & 3600 & 110   & 12.03 & K7  & 4060 &  II & 12.4 \\
BP Tau &   2007-12-20 &  3600 & 110  & 12.16 & K7 & 4060 & II & 7.8 \\
        &             &  1168 & & & & & &\\	
V826 Tau A & 2007-09-19 & 3600 & 90 & 12.11  &  K7 & 4060 & III & 4.2 \\
Hubble 4 & 2007-09-20 & 2x3000  & 120 & 12.67  &  K7 & 4060 & III & 12.8 \\
	&             & 2150    & &   &     &       &  & \\
RXJ0438.2+2023 & 2007-10-30 & 3600  & 110 & 12.22 & K2 & 4900 & III & 8.0\\
	&           & 2550  & &     &  &   & \\
RXJ0423.7+1537 & 2007-09-19 & 2200 & 140  & 11.29 &  K2 & 4900 & III & ... \\
LkCa 19 & 2007-12-20 & 1270  & 100 & 10.85  & K0 & 5250 & III & 18.6 \\
\hline\hline
\end{tabular}
\label{logtau}
\end{center}
\end{table*}
\section{Analysis}\label{analysistau}
\subsection{Estimate of veiling and equivalent-width correction}
Spectral veiling, which often affects spectra of pre-main sequence (PMS) stars, was estimated in the same fashion widely described in 
 D09. 
Briefly, we compared the equivalent widths (EWs) of the
sample stars with those of stars with similar parameters belonging to the older clusters IC~2602 ($\sim$ 30 Myr) and IC~2391 ($\sim$ 50 Myr): the veiling 
parameter is
expressed as 
$r$ = (EW$_{\rm IC*}$/EW$_{\rm TAU}$) $-$1.  

Nine spectral lines, of different atomic species, were employed in this computation, namely Ca~{\sc i} 5857.5 \AA,  Ca~{\sc i} 6102.7 \AA, Ca~{\sc i} 6122.2 \AA, Fe~{\sc i} 6546.3 \AA, 
V~{\sc i} 6624.8 \AA, Ni~{\sc i} 6643.6 \AA, Fe~{\sc i} 6662.5 \AA, Ca~{\sc i} 6717.7 \AA, and Ti~{\sc i} 6743.2 \AA.
In a subsequent step, we then computed for each star the mean $r$ value coming from the different spectral lines;  
these average values, along with the standard deviation, are listed in Table~\ref{restabtau}. 
All the WTTs of our sample have negligible veiling, as expected; on the other hand, the 
two CTTs, GM Aur and BP Tau, exhibit quite large
$r$ values. In particular, our values of 0.23($\pm$0.09) and 0.60($\pm$0.07) are in excellent agreement with the previous 
estimates by Basri et al. (1991), who found 0.25($\pm$0.1) and 0.60($\pm$0.1), respectively, for the two stars.
Finally, the correction for spectral veiling 
was applied to the two ``veiled" stars, by multiplying for $(1+r)$ the observed EWs to obtain the correct values.   
\subsection{Abundance analysis}
LTE abundances were derived by applying the spectral analysis program MOOG of Sneden (1973, 2002 version) and 
using the GAIA set of model atmospheres (Brott \& Hauschildt, private communication)
\footnote{{\url {http://www.hs.uni-hamburg.de/EN/For/ThA/phoenix/index.html}}. 
Biazzo et al. (2010) provide a wide discussion of why this set of model atmospheres were chosen for these 
young, cool stars.}. 
We performed both EW and spectral synthesis analyses, 
using the drivers {\it abfind} and {\it synth}, respectively. 
Radiative, Stark, and collisional broadenings were treated in the standard way, adopting for the last one the 
classical Uns\"{o}ld (1955) approximation. 
\subsubsection{Equivalent width analysis and stellar parameters}\label{ew}
For the five slowest rotators in our sample (vsin$i$$\leq$18 km s$^{-1}$), we derived iron abundances 
by measuring the EWs. The line list was based on the one provided by Randich et al. (2006) to which we added four 
iron lines from the list by D09; we refer to those papers for further details on atomic parameters and, in 
general, for the whole procedure.  

We derived solar abundances by analysing a SARG spectrum of
the Sun taken with the same resolution and grating of our sample stars 
(see Table~\ref{t:solar} for solar abundances 
of Fe, Si, and Ni). 
Our complete line list includes 48, 2, and 13 features of Fe~{\sc i}, 
Si~{\sc i}, and Ni~{\sc i}, respectively; however, depending on the star
temperature and S/N, we were able to measure a different number
of lines for the different stars. The final number of lines used, after
the 2$\sigma$ clipping, is given in Table~\ref{restabtau}.
Assuming T$_{\rm eff}\odot$=5770 K, 
log~g$_{\odot}$=4.44, and $\xi_\odot$=1.1 km s$^{-1}$ (see Randich et al. 2006),
we derived log~n(Fe~{\sc i})$_\odot$=7.51$\pm$0.04. 
The EWs for all the stars were measured using 
the IRAF task {\sc SPLOT} and performing both a Gaussian fit and a direct integration to the line profile.
In the case of Hubble 4, the bluer spectral region (4600~\AA~$-$6200~\AA)
could not be included, since this spectral window was missed because of technical problems during the observations; 
the analysis was thus performed using the upper (i.e., redder) orders 
including only 13 iron lines.
Along with iron, we also derived the abundances of Si~{\sc i} and Ni~{\sc i} for 
the five slow rotators
for which EW analysis could be carried out. 

Initial stellar parameters were retrieved from the XEST source catalogue. 
In particular, we adopted T$_{\rm eff}$ values derived from spectral types and the calibration of KH95. 
 After applying a 2$\sigma$ clipping to the initial line list, we derived the final T$_{\rm eff}$ values
 by zeroing the slope 
 between log~n(Fe~{\sc i}) and excitation potential ($\chi$).
 To infer microturbulence ($\xi$), we first assumed 1.4 km s$^{-1}$ as an initial value, and 
 then imposed the condition of there being no trends between log~n(Fe) and EW strengths.
Surface gravities were computed using the expression 
log~$g$=4.44 +log~(M/M$_{\odot}$)$-$log~(L/L$_{\odot}$)+4$\times$log~T$_{\rm eff}$$-$15.0447, taking mass and luminosity values
from summary tables by G\"{u}del et al. (2007). 
We could not spectroscopically optimize these gravity values, because of the
lack of suitable Fe~{\sc ii} lines in the spectra 
of the cool stars. For the warmest star within the slow-rotating sample, i.e. RXJ0438.2+2023, we were able to measure
only one Fe~{\sc ii} line, lacking statistics. Therefore, photometric gravities are adopted as final values.

In Table~\ref{restabtau}, we list the final stellar parameters: the initial effective temperatures 
agree very well with the final ones, having a maximum difference of 70~K (for BP Tau). 
The final microturbulence values range from 1.3 km s$^{-1}$ for the coolest star BP Tau to 
1.9 km s$^{-1}$ for the warmest one, namely 
RXJ0438.2+2023.  

\setcounter{table}{1}
\begin{table}
\caption{Solar abundances derived from our analysis for Fe, Si, and Ni. As a comparison, we also show in Column 3 the standard abundances by
Anders \& Grevesse (1989).}\label{t:solar}
\begin{center}
\begin{tabular}{lcc}
\hline\hline
Element   &  This work   & AG(1989)\\
\hline    
          &             &  \\
log~n(Si~{\sc i}) & 7.54$\pm$0.03 & 7.55\\   
log~n(Fe~{\sc i}) & 7.51$\pm$0.06 &7.52\\
log~n(Ni~{\sc i}) & 6.24$\pm$0.02 &6.25\\
\hline\hline
\end{tabular}
\end{center}
\end{table}

\subsubsection{Spectral synthesis}
For the two WTT stars with v$sin$i larger than 18 km s$^{-1}$, EW measurements were 
infeasible because of strong line blending.
We thus derived a metallicity estimate using spectral synthesis method in a 
wavelength window of $\sim 15$ \AA, from 6700 \AA~to 6715 \AA, using the line list already employed in D09. 
We first optimised that line list by changing the log$gf$ values, when necessary, to obtain the closest agreement between 
our SARG solar spectrum and the standard iron abundance by Anders \& Grevesse (1989).
To compare the different observed spectra for these rapidly rotating stars with the synthetic spectra, 
we convolved the latter with both a Gaussian profile corresponding to our resolution of R=29000 (FWHM=0.23 \AA) 
and a rotational profile, taking into account the limb-darkening coefficients.

The following stellar parameters were adopted. 
Values of T$_{\rm eff}$ and log~$g$ were retrieved from XEST; we then assumed 
microturbulence of $\xi$=1.6 km s$^{-1}$, since this is the mean 
value obtained from the EW analysis of other sample
stars.
Finally, spectral synthesis allowed us to infer, as a by-product, the projected
rotational velocities v$sin$i 
(Table \ref{restabtau}).

\begin{figure*}
\begin{center}
\includegraphics[width=15cm]{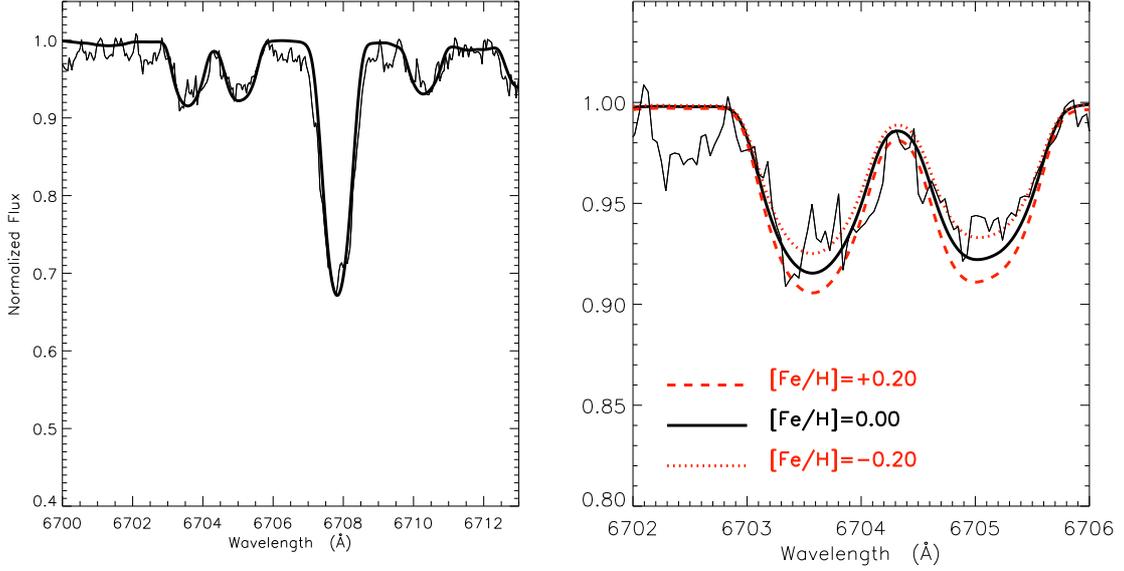}
\caption{Left panel: Best-fit synthetic spectrum ([Fe/H]=0.00) of the star RXJ0423.7+1537. Right panel: Zoom on the small spectral 
window containing 
two Fe~{\sc i} features. Synthetic spectra obtained with the best-fit
[Fe/H] value and its $\pm$ 1 $\sigma$ margins of error superimposed on the observed spectrum 
for comparison (solid, dotted, and dashed lines, respectively).}
\label{synthtau}
\end{center}
\end{figure*}

\begin{figure*}
\begin{center}
\includegraphics[width=15cm]{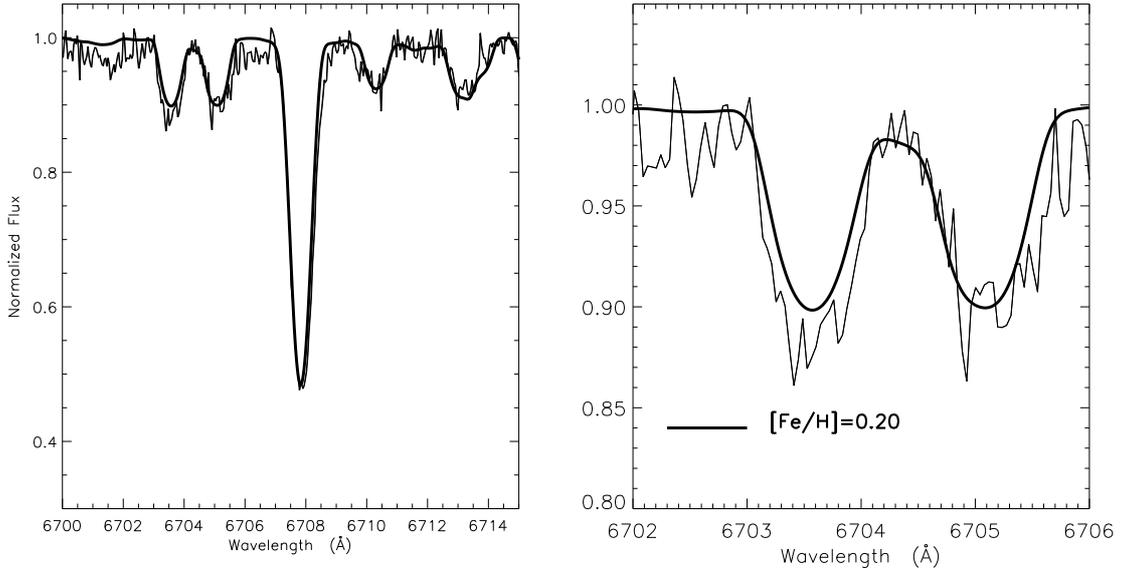}
\caption{Observed and synthetic spectra of the star LkCa19. The ``formal" best-fit is shown
in both panels as a thick solid line
(see text --Section~\ref{sec:comp} for a discussion on this case).}
\label{synth2}
\end{center}
\end{figure*}

In Figures~\ref{synthtau} and \ref{synth2}, examples of spectral synthesis are shown for the stars 
RXJ0423.7+1537 and LkCa19, respectively. They are discussed in Section~\ref{results}.

\subsection{Error budget}
Internal (random) errors, affecting the abundances derived by the EW analysis, 
are due to (i) uncertainties in the oscillator strengths and EW measurements, and (ii)
uncertainties in the adopted set of stellar parameters. 
The first type of uncertainty, i.e., log~$gf$, should be minimised, given
the differential nature of our analysis; on the other hand, uncertainties in EWs are clearly represented by $\sigma$$_1$
in Table~\ref{restabtau}. 
Errors due to uncertainties in T$_{\rm eff}$, log~$g$, and $\xi$  were estimated
as described in D09, to which we refer for a complete and detailed description of the approach followed.
In brief, we assumed typical errors of 60~K in T$_{\rm eff}$, 0.2~dex
in log~$g$, and 0.2 km s$^{-1}$ in $\xi$
since larger errors in T$_{\rm eff}$ and $\xi$ would introduce trends in abundances with $\chi$ and EW,
respectively. For gravity, we instead adopted the same uncertainty as D09. 
We then derived 
the variation in the iron abundance and both [Ni/Fe] and [Si/Fe] by 
changing one stellar parameter at the time, while keeping the others unchanged,
and finally we quadratically added the three contributions
to obtain the total error ($\sigma_2$). 
Furthermore, for the two ``veiled" stars, we computed the abundance error due to the veiling estimate ($\sigma_3$) which implied that, 
by assuming a conservative error 
equal to the rms of the 
mean $r$ value (see Column 2 in Table \ref{restabtau}),    
the variation in the iron abundance is $\pm 0.09$ and $\pm 0.06$ 
for GM Aur and BP Tau, respectively.

Uncertainties related to best-fit determination and stellar parameters affect the abundances derived from spectral synthesis: errors in the best-fit are around 0.1 dex
($\sigma$$_{1}$). Errors in stellar parameters were computed as described by D09: a variation of  
$\pm$60 K in T$_{\rm eff}$ and $\pm$0.2 km s$^{-1}$ in $\xi$, will result in abundance changes of $\pm$0.06 dex;
the effect of log~$g$ is instead smaller, reaching a maximum of $\sim$0.04 dex.  

Systematic (external) errors, caused for instance by the code and/or model atmosphere, should not 
influence largely our final abundances, as widely described in D09. 

\section{Results}\label{results}
\subsection{Mean abundances}\label{sec:mean}
The final abundance values are reported in Table~\ref{restabtau}, for both the EW and spectral synthesis analyses.
For each star (Column 1), we list the veiling parameter $r$ in Column 2, while the 
final (adopted) set of stellar parameters ( T$_{\rm eff}$, 
log$g$, $\xi$)  are 
presented in Columns 3, 4, 5. In Column 6, we show 
the number of lines used in EW analysis (upper panel) and the v$sin$i (km s$^{-1}$) as a by-product of 
spectral synthesis (lower panel);
finally, metallicity [Fe/H] is given in Column 7
along with its corresponding errors. 

Given the difference between the methodologies, we computed two different
average values: the first technique (EW analysis) inferred [Fe/H]=$-$0.02$\pm$0.06, while considering the faster rotator 
RXJ0423.7+1537 (and not considering the less secure
estimate for LkCa 19, see Section~\ref{sec:comp})
we obtain a mean metallicity of 
[Fe/H]=$-$0.01$\pm$0.05. The small standard deviation from the mean, which is much smaller than expected from the formal errors,
indicates a homogeneous solar iron abundance for Tau-Aur with no star-to-star
scatter.

The $\alpha$-element Si~{\sc i} along with the Fe-peak one Ni~{\sc i} also confirm solar values (see Table~\ref{t:xfe}): 
the average ratios [Si/Fe]=0.04$\pm$0.03 and [Ni/Fe]=$-$0.02$\pm$0.05 indicate
that Tau-Aur members are characterised by the
same abundance pattern as thin disc stars of similar metallicity. 

Our study provides the first abundance determination using the EWs of ``deveiled" CTTs, 
following the work of Biazzo et al. (2010) for the Orion complex. A moderately accreting PMS star was also present in the Padgett's 
sample, namely CV Chamaeleonis; however, after the estimate of
spectral veiling (r=0.2$\pm$0.3), the EWs correction was not performed and the author concluded that the derived 
iron abundance should be treated as a lower limit.
We find that CTTs and WTTs share the same chemical 
composition; while this result
indicates that the presence of a circumstellar disc
does not affect the stellar photospheric abundances, it
also strongly supports
the validity of our method for estimating spectral veiling.
\setcounter{table}{2}
\begin{table*}
\begin{center}
\caption{Adopted stellar parameters, number of used Fe lines, and [Fe/H] values with uncertainties.}
\begin{tabular}{lcccccc}
\hline
\hline
star & $r$ & T$_{\rm eff}$ & log~$g$ & $\xi$  &  Nr lines & [Fe/H]$\pm$$\sigma$$_1$$\pm$$\sigma$$_2$$\pm$$\sigma$$_3$\\
     &     &    (K)        &       & (km s$^{-1}$) &   & (EWs)		\\
\hline
           &               &       &      &          &        &\\
GM Aur & 0.23$\pm$0.09(C) & 4100 & 3.9 & 1.7 & 17 &                ~~0.04$\pm$0.08$\pm$0.08$\pm$0.09\\
BP Tau & 0.60$\pm$0.07(C) & 3900 & 3.7 & 1.3 & 18 &               $-$0.07$\pm$0.05$\pm$0.09$\pm$0.06\\
V826 TauA & ~~~~~...~~~~             & 4000 & 3.8 & 1.4 & 28 &                ~~0.00$\pm$0.09$\pm$0.08~~~~~~~~~~\\
Hubble 4 &  ~~~~~...~~~~  & 4000 &  3.3 & 1.5 & ~13$^{\rm a}$   &  $-$0.08$\pm$0.05$\pm$0.08~~~~~~~~~~\\
RXJ0438.2+2023	& ~~~~~...~~~~ &  4850	&  4.5 & 1.9 & 27 &                ~~0.03$\pm$0.09$\pm$0.05~~~~~~~~~~\\
 &       	&    	& 		 &    &    & \\
& & & & & & \\
\hline
          &               &       &               &          &  v$sin$i     & [Fe/H]$\pm$$\sigma$$_1$$\pm$$\sigma$$_2$\\
	  &                &      &              &           &  (km s$^{-1}$)    &  (Synthesis)\\
RXJ0423.7+1537 & ~~~~~...~~~~   & 4900 & 4.3 & 1.6 & 25$\pm$2 & 0.00$\pm$0.20$\pm$0.06\\
&       	&    	& 		 &    &    & \\

{\large AVERAGE}   &    &  &   & &  &  {\large $-$0.01$\pm$0.05}\\

&       	&    	& 		 &    &    & \\
LkCa 19   & ~~~~~...~~~~   & 5250 & 4.2 & 1.6 & 21$\pm$3 & 0.20$\pm$0.20$\pm$0.06\\

 & & &                  &         &           &     \\
\hline
\hline
\end{tabular}
\label{restabtau}
\end{center}
\begin{list}{}{}
\begin{footnotesize}
\item[$^\mathrm{a}$] Only the upper CCD was available
\end{footnotesize}
\end{list}
\end{table*}
\setcounter{table}{3}
\begin{table*}
\begin{center}
\caption{[Si/Fe] and [Ni/Fe] ratios for a sub-sample of our stars}\label{t:xfe}
\begin{tabular}{llclc}
\hline\hline
star      	 &     [Si/Fe]$\pm$$\sigma$$_1$$\pm$$\sigma$$_2$$\pm$$\sigma$$_3$  & Nr lines & 
[Ni/Fe]$\pm$$\sigma$$_1$$\pm$$\sigma$$_2$$\pm$$\sigma$$_3$ & Nr lines\\
\hline 
                 &	                            &	  &                                     &	 \\		     
GM~Aur     	 &  0.04$\pm$0.09$\pm$0.07$\pm$0.03 & 1   &  $-$0.05$\pm$0.04$\pm$0.03$\pm$0.02 &	4 \\ 
BP~Tau     	 &  0.08$\pm$0.07$\pm$0.08$\pm$0.02 & 1   &  ~~0.06$\pm$0.07$\pm$0.04$\pm$0.02  & 5 \\
V826Tau~A  	 &  0.02$\pm$0.02$\pm$0.07          & 1   &  ~~0.01$\pm$0.01$\pm$0.03	      &	10 \\ 
Hubble4   	 &  ~~~~~~~~~~~...~~~~~~~~~~~~~     & ... &  $-$0.02$\pm$0.14$\pm$0.03	      &	2  \\
RXJ0438.2+2023   &  0.03$\pm$0.07$\pm$0.06          & 2   &  $-$0.06$\pm$0.04$\pm$0.03	      &	4  \\
                 &                                  &     &                                     &    \\
{\bf AVERAGE}     &  ~~~~~~~{\bf 0.04$\pm$0.03}     &     &  ~~~~~~~{\bf $-$0.02$\pm$0.05} & \\
\hline	
\end{tabular}
\end{center}
\end{table*}
 \subsection{Comparison with previous works}\label{sec:comp} 
Only one of our stars is in common with previous studies
in the literature: LkCa19. For this star, we obtain a formal best-fit 
with [Fe/H]=+0.2$\pm$0.2 (see Table~\ref{restabtau}), suggesting a possible
iron overabundance. However, the uncertainties in the stellar parameters
of this star prevent us from definitively assessing its metal-rich nature. 
As Figure~\ref{synth2} shows, even the formal 
best-fit synthetic spectrum cannot simultaneously reproduce
both the Fe~{\sc i} lines at 6703 and 6705 \AA, 
suggesting that the adopted T$_{\rm eff}$, as derived from 
spectral-type (T$_{\rm eff}=$5250~K), is likely to be overestimated. We indeed
note that for LkCa19 Martin et al. (1994) derived
a value of T$_{\rm eff}$=4343 K, 
which is probably too low. An intermediate value might be the correct one and
would yield a lower [Fe/H]. A lower effective temperature would also
yield a lower gravity. Since the available spectrum, and in particular the
presence of several blended features, does not allow us to independently
derive stellar parameters, at the present time, no final 
statement can be drawn about the metallicity of this star.

This star was also previously analysed by Padgett (1996), who obtained a 
T$_{\rm eff}$=5214$\pm$146 K, an iron abundance [Fe/H]=$-$0.08$\pm$0.08, and a significantly higher microturbulence
$\xi$=2.8$\pm$0.5 km s$^{-1}$ from an EW analysis. The Padgett's line list contains about 
one third of spectral features with EWs
larger than 150 m\AA~ with a maximum value of 330 m\AA~: these 
high values are probably due to blended features,
given the high rotational velocity of $\sim$20 km s$^{-1}$. In other words, an EW analysis is really 
uncertain in this case, when weak metallic 
lines blend together and the resulting abundance probably has to be revised. 
To conclude, given the uncertainties in effective temperature, hence metallicity, the star LkCa19 was discarded
in the computation of the average abundance for Tau-Aur, as reported in Section~\ref{sec:mean}.

\begin{figure*}
\begin{center}
\includegraphics[width=12cm]{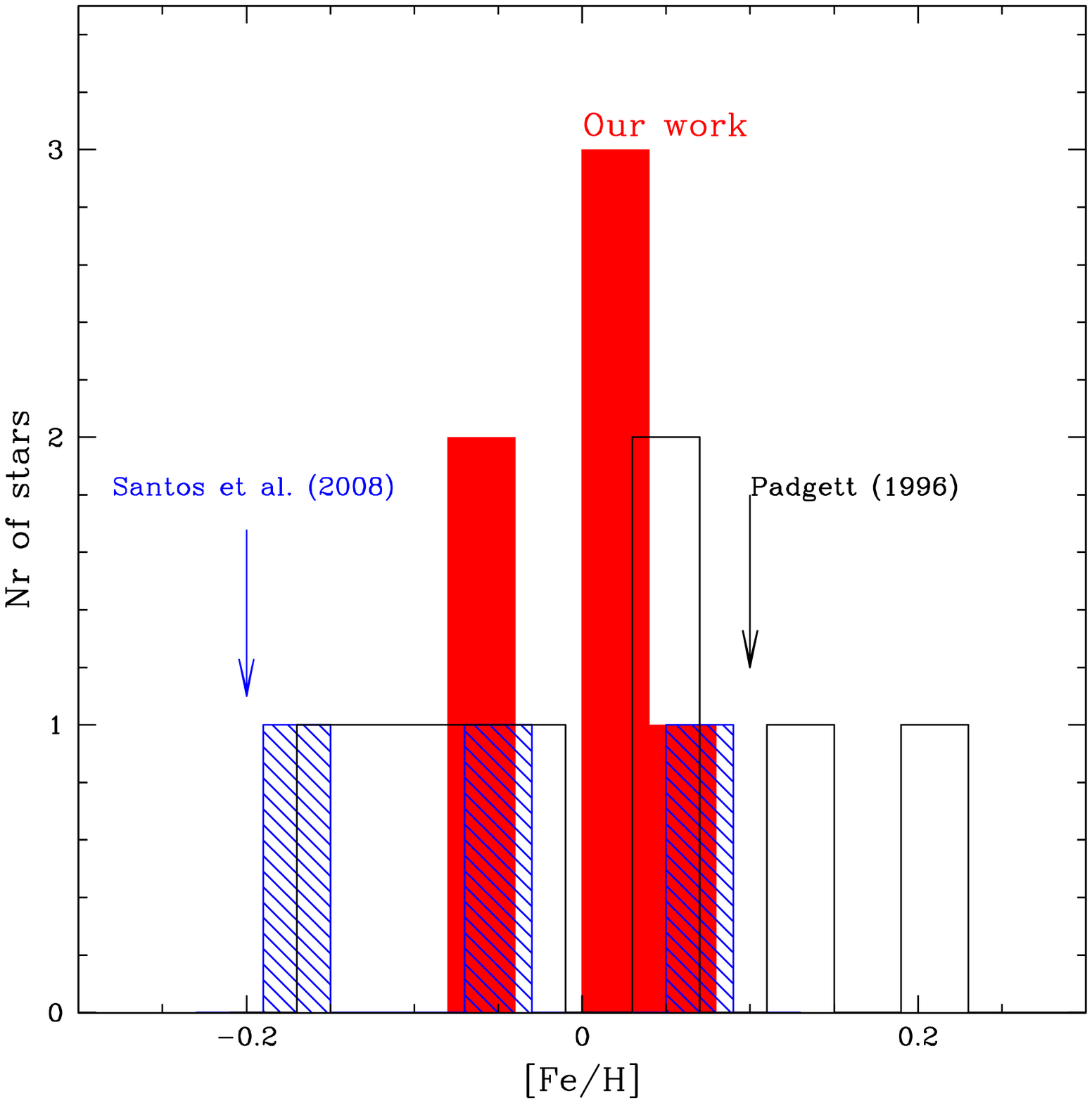}
\caption{Comparison of our [Fe/H] (filled histogram) with previous estimates by Padgett 
(1996, empty histogram) and Santos et al. (2008, dashed histogram).}
\label{comptau}
\end{center}
\end{figure*}

Focusing on the overall metallicity, 
in Figure~\ref{comptau}, our [Fe/H] determination in Tau-Aur is compared with the previous estimates by Padgett (1996) and Santos
et al. (2008); their averaged values of [Fe/H]=$-$0.01$\pm$0.13 and [Fe/H]=$-$0.07$\pm$0.12, respectively, are in good agreement
with our new estimate. 
Moreover, the [Si/Fe]=0.04$\pm$0.04 and [Ni/Fe]=$-$0.07$\pm$0.02 ratios derived by Santos et al. (2008) 
are similar to the values inferred here.

However, the figure clearly shows that the [Fe/H] distribution
from our measurements is much narrower than those
obtained by both those studies. 
As mentioned in Sect.~\ref{introduction}, in the work by  Santos et al. (2008) the metallicity ranges from $-$0.18 to +0.05, while 
in Padgett (1996) the internal dispersion is even larger, with [Fe/H] ranging
from $-$0.16 to +0.22 dex. Padgett concluded that, 
beyond the major uncertainties affecting young stars with respect to the older ones, ``the standard deviation of [Fe/H] 
within the T Tauri star association is larger than older clusters". 
However, our homogeneous abundance determination shows that, at least within our sample, the star-to-star scatter
is smaller than the observational errors and that previously detected 
variations probably reflect these uncertainties. The same conclusion has been drawn by D09 and Biazzo et al. (2010), 
with both studies focusing on the abundance determination across the Orion complex.
On the other hand, we note that the number of SFRs chemically analysed by considering a large sample of stars is still 
very limited; further investigations are necessary to 
confirm (or refuse) the presence of real internal scatter.

\section{Discussion}\label{discu}
\subsection{The metallicity of Taurus-Auriga}
The distribution of [Fe/H] for the few SFRs and young 
associations with available abundance measurements was discussed 
by Santos et al. (2008) and Biazzo et al. (2010). 
None of them is more metal-rich than the Sun and, in general,
they appear to be a part of a rather
homogeneous sample, characterised by similar iron abundance patterns. 
Biazzo et al. obtained an average metallicity [Fe/H]$=-0.06\pm0.04$,
considering SFRs within 500~pc of the Sun and including the
value for Tau-Aur derived in the present paper. 

Nevertheless, we note that small (but larger than the errors) differences
in the metallicity
of the regions are present, with the Orion Nebula Cluster (ONC)
and Chamaeleon being the most metal
poor (with [Fe/H]=$-0.11\pm0.08$ and $-0.11\pm0.11$, respectively)
and Tau-Aur, with a solar metallicity, being the most metal
rich. Although the current sample is still too small to perform 
reliable statistics and draw definite conclusions, it seems that low-mass 
SFRs do not behave differently from high mass SFRs, that metallicity
is not related to the characteristics and position of the region, and that the shape of the IMF is not 
related to the metallicity. In particular,
we notice that a difference in metallicity is most likely
not the reason for the different shapes of the IMF in the Chamaeleon
and Taurus; indeed, as mentioned above, Chamaeleon is 
more metal poor than Taurus
and, accordingly, should produce a smaller number of low-mass stars. We observe instead 
a peak at $\sim 0.8 \rm M_{\odot}$ in the IMF of Taurus, with
a deficit of lower mass stars, at variance with other regions, like Chamaeleon,
showing an IMF peak at lower masses (around 0.1 M$_{\odot}$, see Luhman 2004 and references therein).
\subsection{Coronal abundances}
Scelsi et al. (2007b) determined the coronal abundances of about 20 T Tauri stars in the Taurus Molecular Cloud.
They found a clear inverse-FIP effect, with median Fe and Si abundances
$\sim$0.2 times the solar photospheric abundances.
However, as stressed by Scelsi et al. and in Sect.~1 of this paper, 
the determination of photospheric 
abundances of young (active) stars is critically important to comparing in a homogeneous way these values
with the coronal ones and drawing more definitive conclusions on possible
FIP-related trends.

Three of our stars (BP Tau, V826Tau, and Hubble4) are in common with the 
sample of Scelsi et al. For all three stars,
we obtain solar Fe and Si abundances from our analysis.
This confirms a large depletion of iron and silicon
in the hot coronae of these stars with respect to the photospheres, 
which may be attributed to condensation of iron 
in grains, as suggested by Scelsi et al. (2007b).    
\subsection{Metallicity and planet formation} 
Surveys of old solar-type stars have clearly shown that the probability of hosting giant planets increases 
with the metallicity of the parent star: the frequency of giant planets around stars of twice solar metallicity
is about $\sim$ 30\%, to be compared with $\sim$3\% for stars characterised by solar or sub-solar iron content
(e.g., Santos et al. 2004; Fischer and Valenti 2005). The origin of this correlation between metallicity and planet
incidence is still debated, with the coexistence of different hypotheses, from detectability 
(i.e., the higher the metallicity, the greater the likelihood of planet migration and hence planet detection), 
to environmental effects
(i.e., metals enhances the formation of dust, planetesimals, and finally cores).
We mention in passing that the correlation between
planet and metallicity seems to have two major caveats: (1) 
the different behaviours of dwarfs and giants
(see Pasquini et al. 2007), giant stars hosting planets not
appearing to be more metal-rich on average than star without planets;  and (2)
this correlation is no longer valid for [Fe/H] values ranging from $-$0.7 dex to
$-$0.3 dex (see Haywood 2009). 

The high metal abundances of planet-bearing stars seem to be primordial
(e.g., Pinsonneault et al.~2001;
Ecuvillon et al.~2006; Gilli et al.~2006).
A natural question is therefore whether young PMS stars with 
super-solar metallicity exist, since their circumstellar 
discs are commonly assumed to be the planet birthplace.
As mentioned in Sect.~5.1, all the young clusters/associations for which an abundance
analysis has been performed to date do not contain any high-metallicity star.
For six SFRs, Santos et al. (2008) derived a slightly sub-solar metallicity, concluding that
metal-rich stars in the solar proximity (and especially planet-host stars) might have formed 
in the inner Galaxy and then suffered
radial migration across the Galactic disc. This scenario also seems to be 
confirmed by several studies based on both abundances and kinematics, such as Ecuvillon et al. (2007) and Haywood (2008). 


Given the quite limited sample, no final conclusion can be drawn about the possible presence
of metal-rich SFRs, or at least, a population of metal-rich T Tauri stars inside them. 
However, we can speculate that at the present time in the solar neighbourhood, 
for its young clusters/associations, 
the probability of forming giant planets is not as high as in the inner disc. This does
not mean that no giant planets can form (and possibly survive) at all in these young stellar aggregates, but rather that 
the frequency of giant planets could be somewhat lower than in the inner regions of our Galaxy, where the planet-host stars are 
suggested to originate. 
To explain this, two hypotheses have been proposed: (1) The higher metal content of the inner Galaxy
and thus the higher dust-to-gas ratio can enhance the probability of forming gas giant planets 
(for details of the core accretion model for giant planet
formation see, e.g., Pollack et al. 1996; Ida \& Lin 2004; Mordasini et al. 2009a,b). 
(2) As an alternative, metallicity-independent 
approach, 
Haywood (2009) suggested that giant planet formation could be favoured 
in regions where the density of the molecular hydrogen (which is the primary constituent of these
objects) is significantly higher. We note that the molecular ring of our Galaxy is located 
precisely at the Galactocentric radii of 3-5 kpc where these metal-rich stars are observed (Haywood 2009).
\section{Summary and conclusions}\label{summary}
We have presented elemental abundances of seven stars, both CTTs and WTTs, belonging to the Tau-Aur association.
We have found a very homogeneous, solar metallicity for this T association, deriving 
a mean value of [Fe/H]=$-$0.01$\pm$0.05. Both [Si/Fe] and [Ni/Fe] 
also exhibit solar ratios and agree very well 
with the observed abundance pattern of thin disc stars at the same metallicity.
In contrast to the previous determination of abundances in Tau-Aur,
 which spanned a wide range in [Fe/H], we conclude that the 
 internal dispersion in metallicity for this association is very similar to values derived in older open 
clusters.
Along with our previous project in Orion and the available estimates in the literature, all the SFRs 
surveyed to date seem to share a similar
chemical composition, suggesting a uniform ISM in the solar surroundings at the present time.
In this context, we note that no metal-rich members have been detected 
in all the analysed young associations:
this could validate the idea that metal-rich planet-host stars in the solar circle 
were formed in the inner disc and subsequently moved to their current location.
Their not-so-high metallicity might reflect the rather low frequency of giant planets in the solar neighbourhood, at
variance with that of the inner Galactic regions. 
However, given that only a small number of SFRs have been chemically characterised so far, and a small number of stars per region, 
further investigations are mandatory to definitely help us address this controversial, but intriguing issue. 
\begin{acknowledgements}
 We warmly thank A. Mart{\'i}nez Fiorenzano for his valuable suggestions and helpful comments during the preparation of the observations. 
 This work has made extensive use of WEBDA and SIMBAD databases. The XEST team is kindly acknowledged for having provided the full source catalogue 
 in advance of publication. VD thanks Raffaele Gratton for very useful discussions on metallicity and 
 giant planets. 
 Misha Haywood deserves to be kindly acknowledged for sending us the
 material not directly available from the web.
 Finally, we thank the anonymous referee for her/his valuable suggestions
 and comments who improved the quality of our manuscript.

\end{acknowledgements}

\end{document}